\documentclass{ws-p8-50x6-00}
\usepackage{amsmath}
\usepackage{amssymb}
\usepackage{subfigure}

\begin{document}

\title{Study of Confinement Using the \\Schr\"odinger Functional}

\author{Paolo Cea}

\address{INFN and Dept. of Physics, Univ. of Bari, via Amendola 173,
70126 Bari, Italy\\
E-mail: Paolo.Cea@ba.infn.it}

\author{Leonardo Cosmai}

\address{INFN - Sezione di Bari, via Amendola 173,
70126 Bari, Italy\\
E-mail: Leonardo.Cosmai@ba.infn.it}


\maketitle

\abstracts{
We use a gauge-invariant effective action defined in terms of the
lattice Schr\"odinger functional to investigate vacuum dynamics
and confinement in pure lattice gauge theories. After
a brief introduction to the method, we report some numerical results.
}
\section{Introduction}

To study the vacuum structure of the lattice gauge theories we
introduced~\cite{Cea:1996ff,Cea:1999gn,Cea:an}
a gauge invariant effective action, defined by using the lattice
Schr\"odinger functional.

The Schr\"odinger functional can be expressed as a functional
integral~\cite{Gross:1980br,Rossi:1979jf}
\begin{equation}
\label{Zetaint} {\mathcal{Z}}[\mathbf{A}^{(f)},\mathbf{A}^{(i)}] =
\int {\mathcal{D}}A_\mu \; e^{-\int_0^T dx_4 \, \int d^3\vec{x} \,
{\mathcal{L}}_{YM}(x)}  \,,
\end{equation}
with the
constraints $\mathbf{A}(x_0=0) = \mathbf{A}^{(i)}$,
$\mathbf{A}(x_0=T) =  \mathbf{A}^{(f)}$,
where $\mathbf{A}(\vec{x})$ are static classical gauge fields. 
The Schr\"odinger
functional Eq.~(\ref{Zetaint}) is invariant under arbitrary static gauge
transformations of $\mathbf{A}(\vec{x})$'s fields.
The lattice implementation of the Schr\"odinger functional 
is discussed in Ref.~\cite{Luscher:1992an}. 

Our lattice effective action for 
the static background field $\mathbf{A}^{\mathrm{ext}}(\vec{x}) =
\mathbf{A}^{\mathrm{ext}}_a(\vec{x}) \lambda_a/2$ 
($\lambda_a/2$ generators of the SU(N) algebra) is defined as
\begin{equation}
\label{Gamma}
 \Gamma[\mathbf{A}^{\mathrm{ext}}] = -\frac{1}{T} \ln
\left\{ \frac{{\mathcal{Z}}[U^{\mathrm{ext}}]}{{\mathcal{Z}}[0]}
\right\} \,\,,\,\,\,\, \mathcal{Z}[U^{\mathrm{ext}}_\mu] =
\int_{U_k(x)|_{x_4=0} = U^{\mathrm{ext}}_k(x)} {\mathcal{D}}U \;
e^{-S_W} \,.
\end{equation}
$\mathcal{Z}[U^{\mathrm{ext}}_\mu]$ is the lattice Schr\"odinger
functional (invariant, by definition, for lattice gauge
transformations of the external links), $U^{\mathrm{ext}}_\mu(x)$ is
the lattice version of the external continuum gauge field
$\mathbf{A}^{\mathrm{ext}}(x)$, and $S_W$ is the standard Wilson action.
$\mathcal{Z}[0]$ is the lattice Schr\"odinger functional with
$\mathbf{A}^{\mathrm{ext}}=0$ ($U^{\mathrm{ext}}_\mu={\mathbf{1}}$).

Our definition of lattice effective action can be extended to
gauge systems at finite temperature as
\begin{equation}
\label{ZetaTnew}
\mathcal{Z}_T \left[ \mathbf{A}^{\text{ext}}
\right] =
\int_{U_k(\beta_T,\vec{x})=U_k(0,\vec{x})=U^{\text{ext}}_k(\vec{x})}
\mathcal{D}U \, e^{-S_W}  \,, \,\,  \beta_t=L_4=\frac{1}{a T} \,.
\end{equation}
The integrations are over the dynamical links with
periodic boundary conditions  in the time direction.
If we send the
physical temperature to zero 
the thermal functional
Eq.~(\ref{ZetaTnew}) reduces to the zero-temperature
Schr\"odinger functional.

\section{Abelian Monopoles and  Vortices}

Monopole or vortex condensation can be detected by means of a disorder parameter $\mu$ 
defined in terms of the lattice Schr\"odinger functional $\mathcal{Z}[ \mathbf{A}^{\text{ext}}]$
introduced in the previous Section.
At zero-temperature
\begin{equation}
\label{disorder}
\mu = e^{-E_{\text{b.f.}} L_4} = \frac{\mathcal{Z} \left[
\mathbf{A}^{\text{ext}} \right]}{\mathcal{Z}[0]} \,,
\end{equation}
$\mathbf{A}^{\text{ext}}$ is the monopole or vortex static background
field. According to the physical
interpretation of the effective action Eq.~(\ref{Gamma})
$E_{\text{b.f.}}$ is the energy to create a monopole or a vortex
in the quantum vacuum. If there is condensation, then
$E_{\text{b.f.}}=0$ and $\mu = 1$.

At finite temperature the disorder parameter is defined in terms of the thermal
partition function Eq.~(\ref{ZetaTnew}) in presence of the given static background
field 
\begin{equation}
\label{disorderT}
\mu = e^{-F_{\text{b.f.}}/T_{\text{phys}}} = \frac{\mathcal{Z}_T
\left[ \mathbf{A}^{\text{ext}} \right]} {\mathcal{Z}_T[0]} \,,
\end{equation}
$F_{\text{b.f.}}$
is now the free energy to create a monopole or a vortex (if there is
condensation $F_{\text{b.f.}}=0$  and $\mu = 1$).

Our disorder parameter $\mu$  is invariant for
time-independent gauge transformations of the external background
fields. This implies that we have not to fix the gauge before
performing the Abelian projection.
Indeed, after choosing the Abelian
direction, needed to define the Abelian monopole or vortex fields
through the Abelian projection, due to
gauge invariance of
the Schr\"odinger functional for transformations of background field,
our results do not depend on the selected Abelian direction,
which, actually, can be varied by a gauge transformation.

\subsection{U(1)  monopoles and vortices}

In the U(1) l.g.t. we considered a Dirac magnetic monopole
background field. In the continuum 
\begin{equation}
\label{monopu1}
e \vec{b}({\vec{x}}) =  \frac{n_{\mathrm{mon}}}{2} \frac{ \vec{x}
\times \vec{n}}{|\vec{x}|(|\vec{x}| - \vec{x}\cdot\vec{n})} \,,
\end{equation}
$\vec{n}$ is the direction of the Dirac string,  $e$
is the electric charge and, according to the Dirac quantization condition,
$n_{\mathrm{mon}}$ is an integer
(magnetic charge = $n_{\mathrm{mon}}/2e$).
The lattice implementation of the continuum field Eq.~(\ref{monopu1}) is straightforward.
As well we can consider a vortex background field:
\begin{equation}
\label{u1vort}
A^{\text{ext}}_{1,2} = \mp \frac{n_{\text{vort}}}{e} \frac{x_{2,1}}{(x_1)^2
+ (x_2)^2} \,,  \quad  A^{\text{ext}}_3 =0  \,.
\end{equation}
We can evaluate, by lattice numerical simulation, the energy to create a Dirac monopole
or a vortex. It is easier to first evaluate the derivative 
$E^\prime_{\mathrm{mon}} = \partial E_{\mathrm{mon}} /
\partial \beta$ ($\beta=1/g$, $g$ is the gauge coupling constant):
\begin{equation}
\label{avplaqu1}
E^\prime_{\text{mon,vort}} = V \left[ <U_{\mu\nu}>_{n_{\text{mon,vort}}=0} -
<U_{\mu\nu}>_{n_{\text{mon,vort}} \ne 0} \right] \,,
\end{equation}
$V$ is the lattice spatial volume. $E_{\text{mon,vort}}$ is then computed by means of a 
numerical integration in $\beta$. 
Our numerical results~\cite{Cea:an} show that Dirac monopoles condense in 
the confined phase (i.e. for $\beta \lesssim 1.01$) of U(1) lattice gauge theory.
While in the case of vortices we do not find a signal of condensation.

Thus, we may conclude that in U(1) lattice theory the strong
coupling confined phase is intimately related to magnetic monopole
condensation~\cite{Fradkin:1978th}.

\subsection{SU(2)  Abelian  monopoles and Abelian vortices}

It is well known that SU(2) lattice gauge theory at finite temperature undergoes a 
transition between confined and deconfined phase.
We studied if Abelian monopoles or Abelian vortices condense in the confined phase
of SU(2). To this purpose we considered in turn an Abelian monopole and an 
Abelian vortex background field. We found~\cite{Cea:an} that both Abelian monopoles and
Abelian vortices condense in the confined phase of SU(2).

\subsection{SU(3)  Abelian  monopoles and Abelian vortices}

For SU(3) gauge theory the maximal Abelian group is
U(1)$\times$U(1), therefore we may introduce two independent types
of Abelian monopoles or Abelian vortices associated respectively to
the $\lambda_3$ and the $\lambda_8$ diagonal generator (one can also consider~\cite{Cea:an}
linear combinations of $\lambda_3$ and  $\lambda_8$). 

Let us focus on the $\lambda_8$ Abelian monopole 
($T_8$ monopole):
\begin{equation}
\label{t8linkssu3m}
U_{1,2}^{\text{ext}}(\vec{x}) =
\begin{bmatrix}
e^{i \theta^{\text{mon}}_{1,2}(\vec{x})} & 0 & 0 \\ 0 &  e^{i
\theta^{\text{mon}}_{1,2}(\vec{x})} & 0 \\ 0 & 0 & e^{- 2 i
\theta^{\text{mon}}_{1,2}(\vec{x})}
\end{bmatrix}
\,, \quad U^{\text{ext}}_{3}(\vec{x}) = {\mathbf 1} \;  \, ,
\end{equation}
with
\begin{equation}
\label{thetat8su3m}
\theta^{\text{mon}}_{1,2}(\vec{x})  = \frac{1}{\sqrt{3}} \left[
 \mp  \frac{n_{\text{mon}}}{4}
\frac{(x_{2,1}-X_{2,1})}{|\vec{x}_{\text{mon}}|}
\frac{1}{|\vec{x}_{\text{mon}}| - (x_3-X_3)} \right] \,.
\end{equation}
Analogously, we can define the $T_3$ Abelian vortex.

Fig.~1 shows that both $T_8$ Abelian monopoles and Abelian vortices condense
in the confined phase of SU(3) l.g.t. at finite temperature
(simulations have been performed on $32^3 \times 4$ lattice using the APE100 crate in Bari).
\vspace{-0.5cm}
%
\begin{figure}[!ht]
\centering
\mbox{\hspace{-0.5cm}
\subfigure[]{\includegraphics[clip,width=0.5\textwidth]{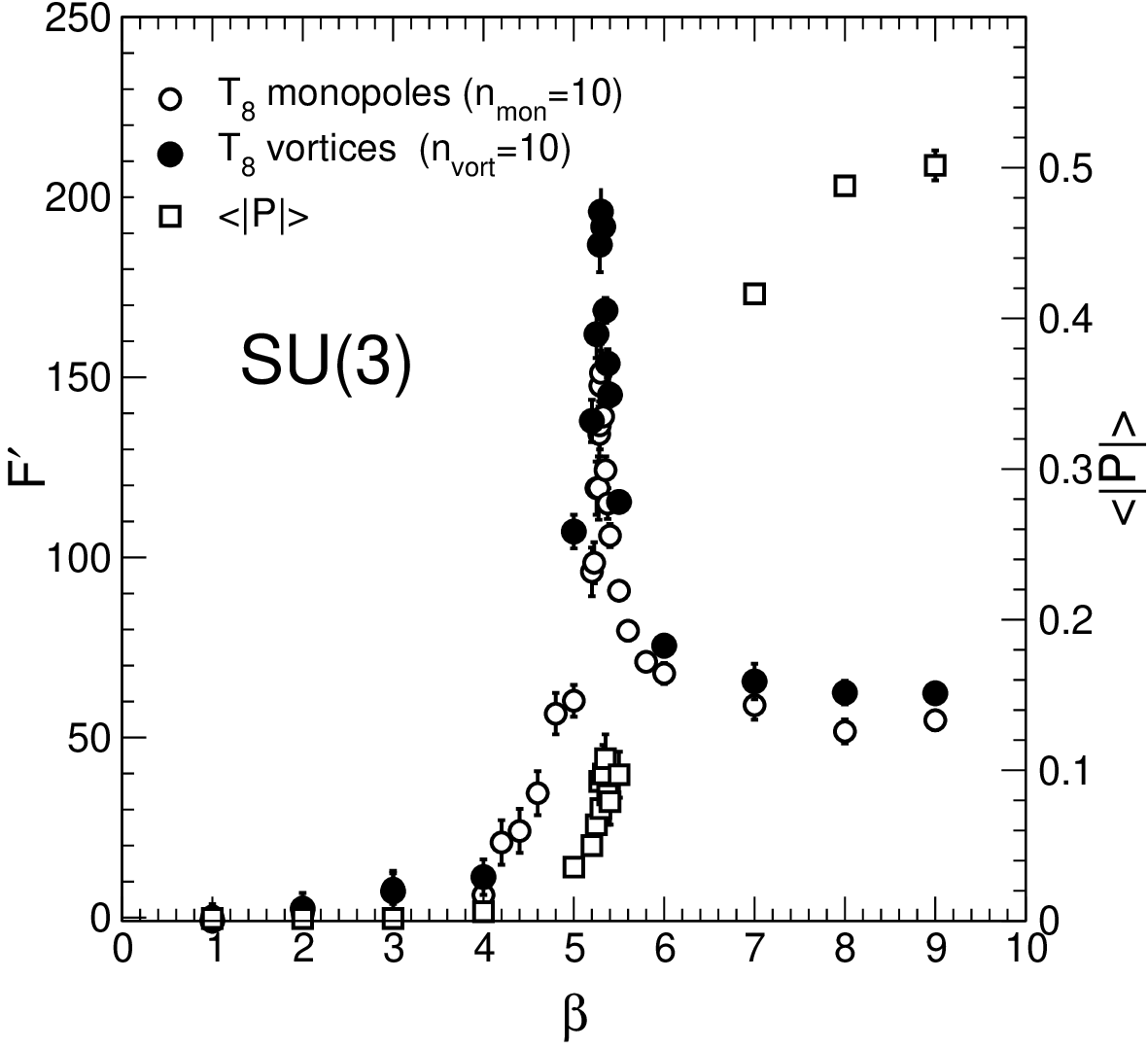}}\hspace{0.5cm}
\subfigure[]{\includegraphics[clip,width=0.5\textwidth]{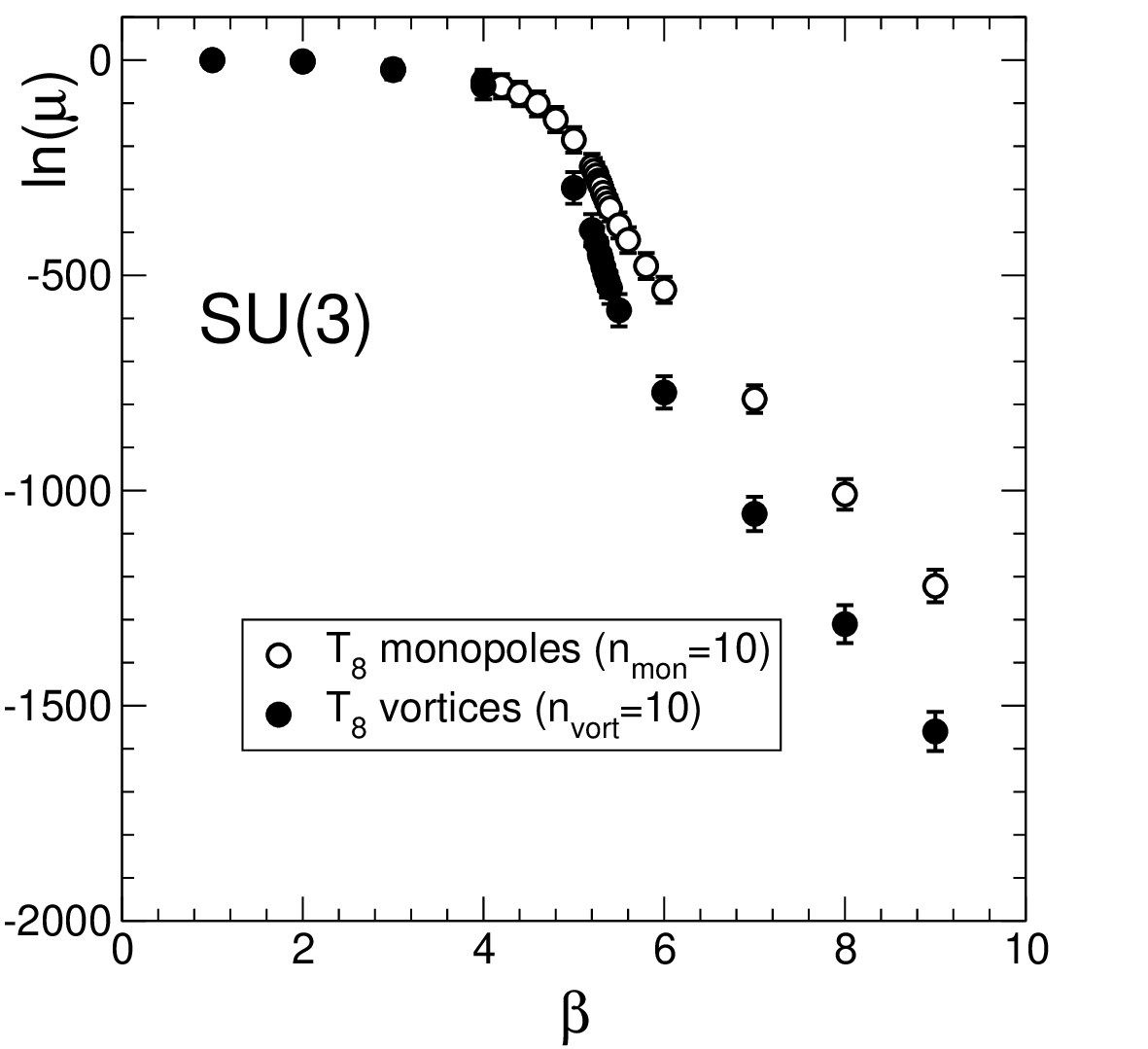}}
}
\vspace{-0.7cm}
\caption{(a) The derivative of the free energy versus $\beta$ for monopoles (open
circles) and vortices (full circles). The absolute value of the Polyakov loop
is also displayed (open squares).
(b) The logarithm of the disorder parameter, Eq.~(\ref{disorderT}), versus $\beta$
for $T_8$ monopoles (open circles) and $T_8$ vortices (full circles).}
\end{figure}
%
\vspace{-0.5cm}
\subsection{SU(3) Center Vortices}

In the case of center vortices the thermal partition function
$\mathcal{Z}_T[{\mathcal{P}_{\mu \nu}}]$ is defined~\cite{Kovacs:2000sy,DelDebbio:2000cx}
by multiplying by the center element $\exp(i 2 \pi/3)$ the set $\mathcal{P}_{\mu
\nu}$ of plaquettes
$\mathcal{P}_{\mu \nu}(x_1,x_2,x_3,x_4)$ with $(\mu,\nu)=(4,2)$,
$x_4=x_4^\star$, $x_2=\frac{L_s}{2}$ and $L_s^{\text{min}} \le
x_{1,3} \le L_s^{\text{max}}$ , with $L_s$ the lattice spatial
linear size.
By numerical integration of
$F^{\prime}_{\text{vort}}$ we can compute $F_{\text{vort}}$ and
the disorder parameter $\mu$ (see Eq.~(\ref{disorderT})). Our
numerical results (see Fig.~2) suggest that in the confined phase 
$F_{\text{vort}}=0$ (in the thermodynamic
limit) and center vortices condense.

%
%
%
%
%
\begin{figure}[htb]
\begin{minipage}[t]{60mm}
\includegraphics[width=1.0\textwidth,clip]{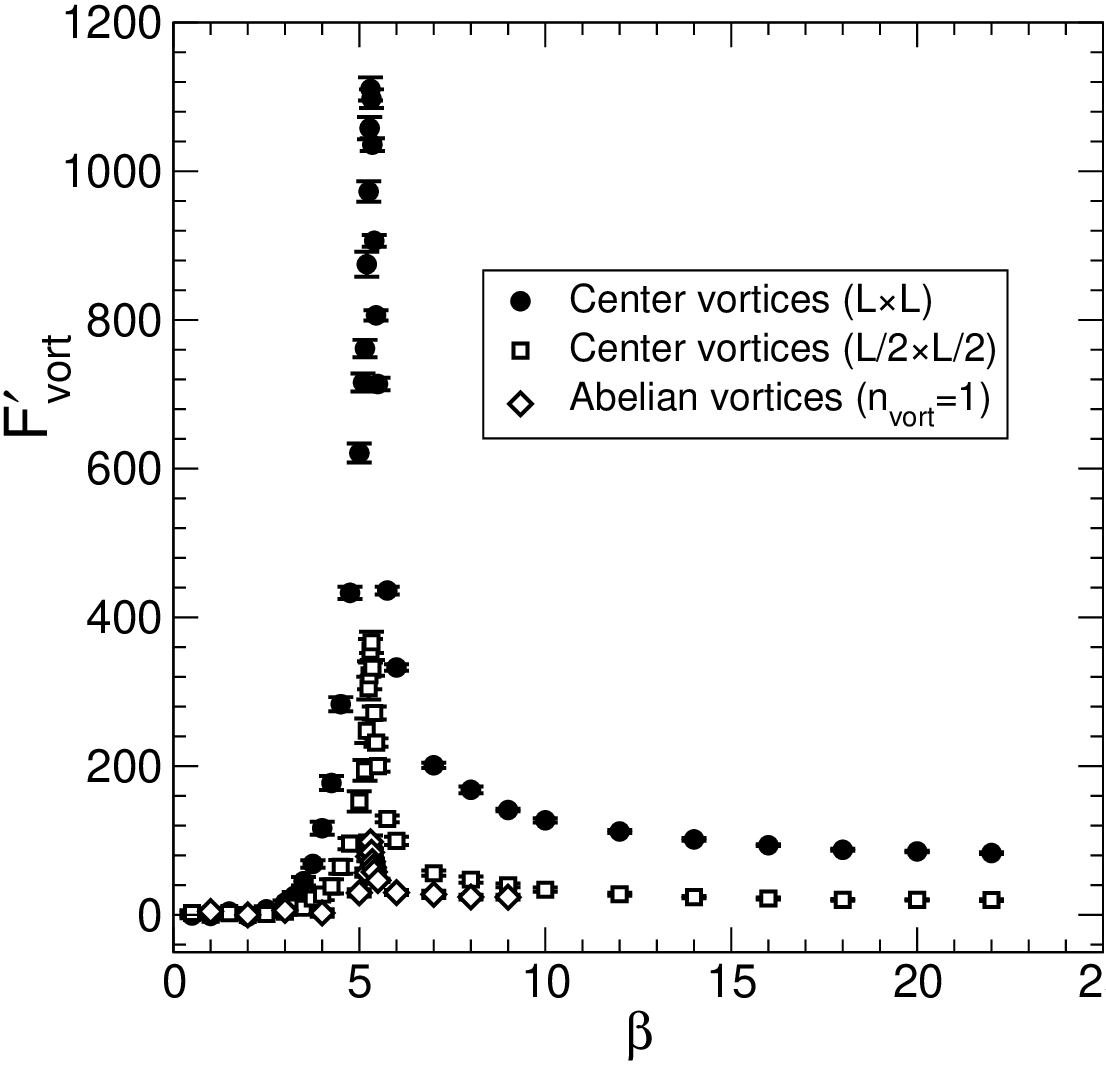}
\vspace{-0.5cm} \caption{$F^{\prime}_{\text{vort}}$ for center
vortices  and $T_8$ Abelian vortices
(vortex charge $n_{\text{vort}}=1$).}
\label{fig:2}
\end{minipage}
\hspace{\fill}
\begin{minipage}[t]{60mm}
\includegraphics[width=1.0\textwidth,clip]{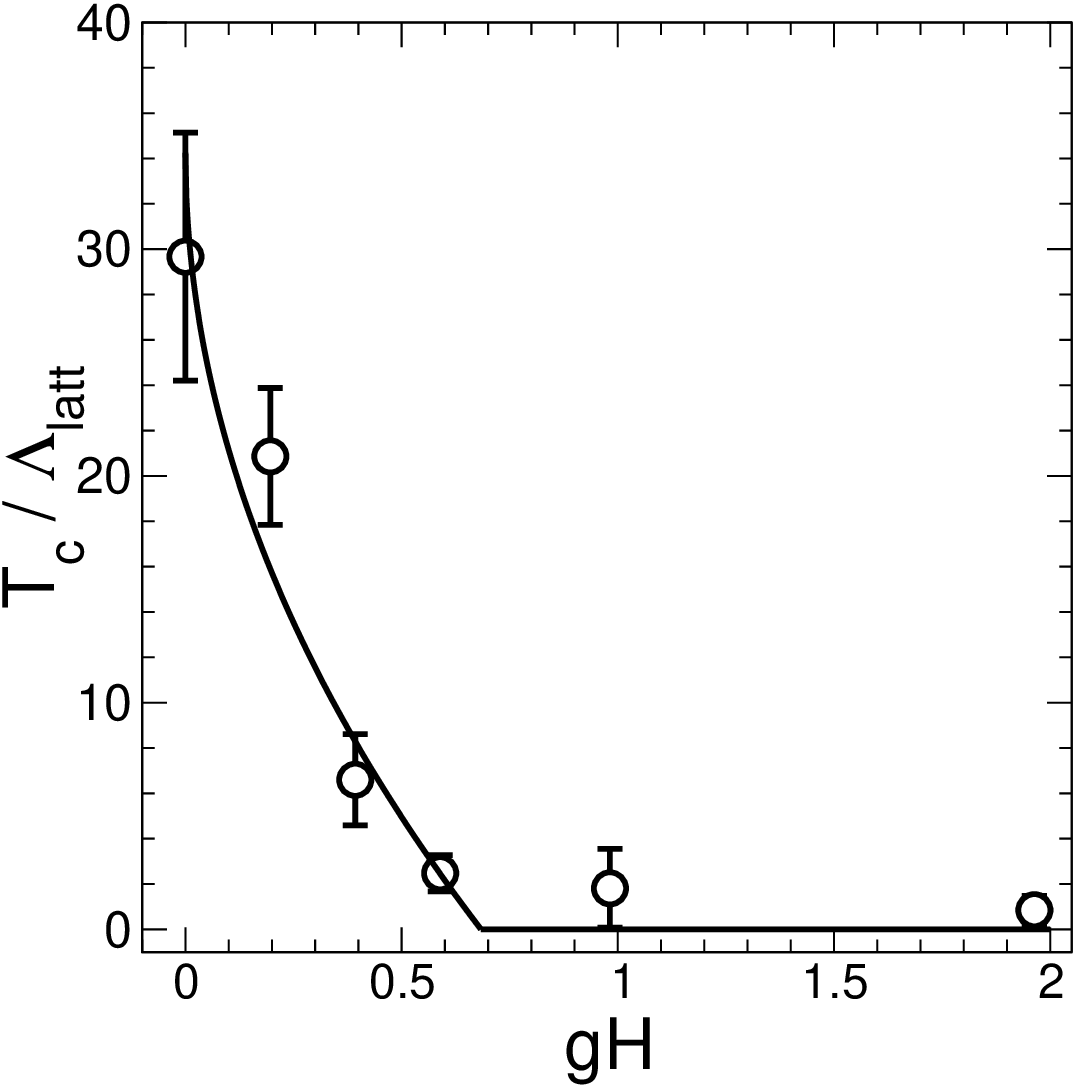}
\vspace{-0.5cm} \caption{$T_c/\Lambda_{\text{latt}}$ versus the
applied external field strength $gH$.} \label{fig:3}
\end{minipage}
\end{figure}
%

\vspace{-0.3cm}
\section{Constant Abelian Chromomagnetic Field}

We want to study the SU(3) gauge system at finite temperature in presence of
an external constant Abelian magnetic field
\begin{equation}
\label{field}
\vec{A}^{\mathrm{ext}}_a(\vec{x}) =  \vec{A}^{\mathrm{ext}}(\vec{x}) \delta_{a,3} \,, \quad
\vec{A}^{\mathrm{ext}}_k(\vec{x}) =  \delta_{k,2} x_1 H \,.
\end{equation}
Spatial links  belonging to a given time slice are fixed to 
\begin{equation}
\label{t3links}
U^{\mathrm{ext}}_1(\vec{x}) = U^{\mathrm{ext}}_3(\vec{x}) =
 {\mathbf{1}} \,, \quad
U^{\mathrm{ext}}_2(\vec{x}) = 
\begin{bmatrix}
e^{i \frac {g H x_1} {2} } & 0 & 0 \\ 0 &  e^{- i \frac {g H x_1}{2}}  & 0
\\ 0 & 0 & 1
\end{bmatrix}
\; \, ,
\end{equation}
that corresponds to the continuum gauge field in Eq.~(\ref{field}).
The magnetic field $H$ turns out to be quantized (due to periodic boundary conditions): 
$a^2 g H/2 = (2 \pi)/ L_1) n_{\text{ext}}$ ($n_{\mathrm{ext}}$ integer).

Since the gauge potential in Eq.~(\ref{field}) gives rise to a constant field
strength we can consider the density $f[\vec{A}^{\mathrm{ext}}]$ 
of the free energy functional $F[\vec{A}^{\mathrm{ext}}]$ 
\begin{equation}
\label{free-energy}
f[\vec{A}^{\mathrm{ext}}] = \frac{1}{V} F[\vec{A}^{\mathrm{ext}}] = - \frac{1}{V L_t}
\ln \frac{\mathcal{Z_T}[\vec{A}^{\mathrm{ext}}]}
{{\mathcal{Z_T}}[0]} \,, \quad V=L_s^3 \,.
\end{equation}
As is well known, the pure gauge system undergoes
the deconfinement phase transition by increasing the temperature.
The deconfinement temperature in $\Lambda_{\mathrm{latt}}$ units is
\begin{equation}
\label{Tc}
\frac{T_c}{\Lambda_{\mathrm{latt}}} = \frac{1}{L_t}
\frac{1}{f_{SU(3)}(\beta^*(L_t))} \,,
\end{equation}
where $f_{SU(3)}(\beta)$ is the two-loop asymptotic scaling function and 
$\beta^*(L_t)$ is the pseudocritical coupling $\beta^*(L_t)$ at a given
temporal size $L_t$, and can be determined by fitting the peak of
$f^\prime[\vec{A}^{\mathrm{ext}}]=\partial f[\vec{A}^{\mathrm{ext}}]/\partial \beta$ 
for the given $L_t$.

Following~\cite{Fingberg:1992ju}
we can perform a linear extrapolation to the continuum of our data
for $T_c/\Lambda_{\mathrm{latt}}$.
We vary the strength of the applied external
Abelian chromomagnetic background field in order to analyze
 a possible dependence of
$T_c$ on $gH$.
We perform  numerical simulations on $64^3 \times L_t$ lattices
with $n_{\text{ext}}=1,2,3,5,10$.
Our numerical results show that the critical temperature decreases
by increasing the external Abelian chromomagnetic field.
For dimensional reasons one expects that
$T_c^2 \, \, \sim \,  \, gH \,$ .
Indeed we get a satisfying fit to  our data with
\begin{equation}
\label{Tcfit}
\frac{T_c(gH)}{\Lambda_{\text{latt}}} = t + \alpha \sqrt{gH} \,.
\end{equation}
From Fig.~3 one can see that there exists a critical field $H_c$ such that
the deconfinement temperature $T_c=0$ for $H>H_c$.

\end{document}